\documentclass[11pt,a4paper]{article}
\usepackage[hyperref]{emnlp2020}
\usepackage{amsmath}
\usepackage{fix-cm}
\usepackage{graphicx}
\usepackage{latexsym}
\usepackage{lmodern}
\usepackage{svg}
\usepackage{times}

\usepackage{microtype}

\aclfinalcopy 

\newcommand*\samethanks[1][\value{footnote}]{\footnotemark[#1]}

\title{TextAttack: Lessons learned in designing Python frameworks for NLP}

\author{
    John X. Morris\thanks{~ equal authorship}~~~~~Jin Yong Yoo\samethanks~~~~~Yanjun Qi \\
    University of Virginia \\
    \texttt{\{jm8wx, jy2ma, yq2h\}@virginia.edu}
}

\date{}

\begin{document}
\maketitle

\begin{abstract}
TextAttack is an open-source Python toolkit for adversarial attacks, adversarial training, and data augmentation in NLP. TextAttack unites 15+ papers from the NLP adversarial attack literature into a single framework, with many components reused across attacks. This framework allows both researchers and developers to test and study the weaknesses of their NLP models. To build such an open-source NLP toolkit requires solving some common problems: How do we enable users to supply models from different deep learning frameworks? How can we build tools to support as many different datasets as possible? We share our insights into developing a well-written, well-documented NLP Python framework in hope that they can aid future development of similar packages.
\end{abstract}

\section{Introduction}

Deep neural network (DNN) models have seen dominant use in NLP tasks such text classification, natural language inference, machine translation, and question answering. However, despite their state-of-the-art performance, NLP DNNs are still vulnerable to adversarial attacks \cite{Survey-NLPAE-Zhang2020-cs}. As a result, there have been growing efforts to develop tools that can help researchers and developers better understand the capability of their NLP models. Both \citet{wallace2019allennlp} and \citet{tenney2020language} introduced web-based visual interactive tools that enable users to see model's local explanations. \citet{checklist-2020} introduced a behavioral testing framework that runs a suite of tests to sanity check NLP models. 

One of the challenges for building such tools is that the tool should be flexible enough to work with many different deep learning frameworks (e.g. PyTorch, Tensorflow, Scikit-learn). Also, the tool should be able to work with datasets from various sources and in various formats. Lastly, the tools needs to be compatible with different hardware setups. 

We developed TextAttack, an open-source Python framework for adversarial attacks, adversarial training, and data augmentation. Our modular and extendable design allows us to reuse many components to offer 15+ different adversarial attack methods proposed by literature. Our model-agnostic and dataset-agnostic design allows users to easily run adversarial attacks against their own models built using any deep learning framework.


\begin{figure}[t]
    \centering
    \includegraphics[width=\linewidth]{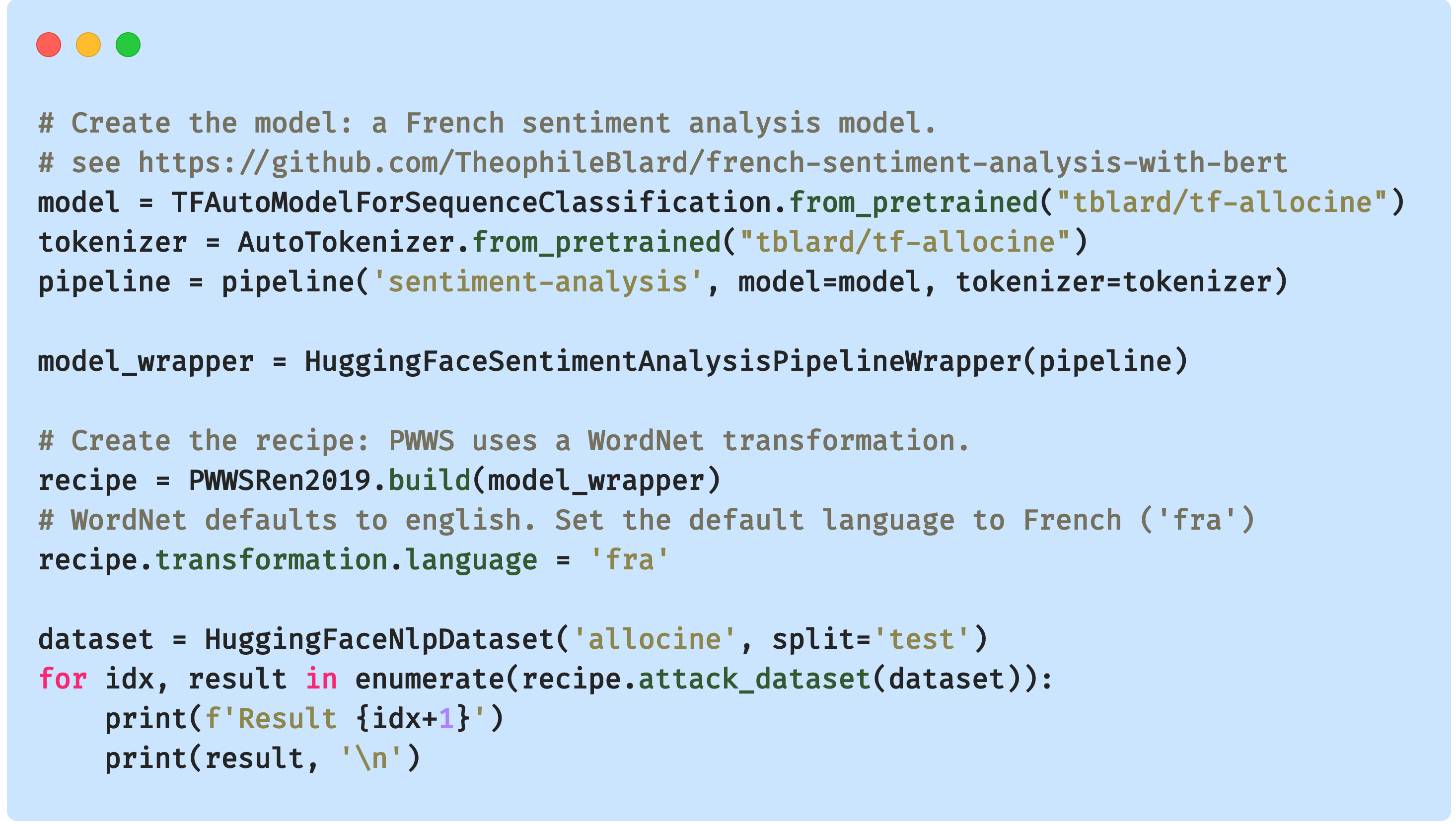}
    \caption{Example usage of the TextAttack API. CamemBERT \cite{CamemBERT-Martin2019-vd} and its tokenizer are initialized using HuggingFace transformers \cite{transformers-Wolf2019-uy} and wrapped in TextAttack model wrappers. Adversarial attack is PWWS \cite{CamemBERT-Martin2019-vd} modified to use WordNet in French \cite{French-WordNet} instead of English. TextAttack's flexible API makes these customizations possible in just a few lines of code.}
    \label{fig:camembert}
\end{figure}

This paper describes some lessons learned along the path to creating TextAttack. Figure \ref{fig:camembert} shows our API in action. Our advice is tailored towards researchers developing NLP libraries in Python that support a variety of models and datasets, and use them for downstream applications.

We provide the following broad advice to help other future developers create user-friendly NLP libraries in Python:
\begin{enumerate}
    \itemsep 0em 
    \vspace{-0.25em}
    \item To become model-agnostic, implement a model wrapper class.
    \item To become data-agnostic, take dataset inputs as (input, output) pairs, where each model input is represented as an \texttt{OrderedDict}.
    \item Do not plan for inputs (tensors, lists, etc.) to be a certain size or shape unless explicitly necessary.
    \item Centralize common text operations, like parsing and string-level operations, in one class.
    \item Whenever possible, cache repeated computations, including model inferences.
    \item If your program runs on a single GPU, but your system contains $N$ GPUs, you can obtain an performance boost proportional to $N$ through parallelism.
    \item Dynamically choose between devices. (Do not require a GPU or TPU if one is not necessary.)
\end{enumerate}

\section{Model agnosticism}
There are growing number of deep learning frameworks and different researchers and groups have preferences about which frameworks to use for different tasks. Unless the library relates to model training or development (and sometimes then), it is possible to build a library that supports deep learning models from any framework.

TextAttack supports both black-box and white-box attacks on NLP models. Black-box attacks can only access the model for inference. In essence, the attack sends lists of text to the model and receives predictions. Model predictions come as lists of floats (for classification), strings, or dictionaries. No other information about the model is required. From the start, we wanted TextAttack to work on models from any framework, without too much headache.

\subsection{Original approach: ``magic'' (model detection logic)}
Our original approach was to take a model and tokenizer as input to each attack and wrangle data into the correct format behind the scenes. This involved a complex series of decisions based by checking the format of the dataset, testing model and tokenizer superclasses, and handling errors as they arose. In the end, it worked: based on the model, tokenizer, and dataset, as well as based on errors raised by passing different data formats to the model, we could perform inference on PyTorch and TensorFlow models. It was ugly, but it worked.

This approach did not scale as there were many edge cases. For example, some TensorFlow Hub models were designed to take strings as predictions, and did not have a tokenizer at all. Some Scikit-learn models took a dataframe as input. We supported both these use cases, but edges cases requiring complex workarounds kept popping up, with no clear end in sight.

\subsection{Better approach: model wrappers}

Our long-term solution was to abstract away the tokenizer and require a new \textbf{model wrapper} class for each model. The idea of \textbf{model wrappers} is that each model is wrapped in a model wrapper that implements a single function, \texttt{\_\_call\_\_}, which takes a list of text inputs and returns a list of predictions. We designed TextAttack to interact exclusively with the model wrapper– not directly with the model, or the tokenizer. 

Model wrappers allow each model to handle its own internals: including tokenization and batch size. TextAttack does not know or care about how information is tokenized before it's sent to the model. TextAttack sends the model a list of strings and receives a \texttt{list}, \texttt{numpy.ndarray}, or \texttt{torch.Tensor} of predictions. 

In this way, TextAttack becomes totally model-agnostic: any user can implement a model wrapper to enable compatibility for a new model or framework. To make the process easier, TextAttack provides model wrappers for common frameworks and patterns. Currently, TextAttack provides model wrappers and example for models implemented with PyTorch \cite{pytorch-Paszke2019-om}, HuggingFace transformers \cite{transformers-Wolf2019-uy}, TensorFlow \cite{tensorflow-Abadi2016-se}, Scikit-learn \cite{scikit-Pedregosa2011-sm}, and AllenNLP \cite{allennlp-Gardner2018-qp}.

\section{Data agnosticism}

Another goal of TextAttack was to be able to run the same attack on any dataset. This has obvious benefits: two attacks that report results on different datasets can easily be compared with TextAttack.

\subsection{Text inputs as \texttt{OrderedDict} objects}
We rely on other libraries for providing default datasets. We provide dataset wrappers for loading datasets from these external libraries. We also allow users to provide their own datasets– via CSV files or Python scripts that load datasets. In essence, each dataset is a list of \texttt{(input, output)} pairs. Each text input is a string (for single-input tasks) or an \texttt{OrderedDict} (for tasks that require more complex input formats).

Each input is an \texttt{OrderedDict} for two reasons: (i) to maintain column labels for display purposes and to make column-specific logic possible and (ii) to maintain ordering so that inputs can be provided to the model in the proper order. An individual text input to the model is a tuple of strings. 

To create these \texttt{OrderedDict} objects from dictionaries loaded from popular dataset libraries, we maintain a tuple of input columns and a string representing the output column. Then, objects from any dataset can be mapped to a data pair for TextAttack: the input is an OrderedDict created from taking the input values in order of the input columns, and an output is the value corresponding to the dataset's output column.

\section{Model output flexibility with \texttt{GoalFunction}}
With the proper input and ouput columns and a corresponding model, adversarial attacks can be run on any dataset on any model. Models may have different output formats. For example, a sentiment classifier would produces a list of the probabilities of each class, while a sequence-to-sequence models produce a text output. Task-specific subclasses of the TextAttack \texttt{GoalFunction} class allow adversarial attack goal functions to be defined at a high level, such that the same goal function can be used for any model with the same output type. For example, the \texttt{MinimizeBleuScore} goal function attempts to minimize the BLEU score \cite{BLEU-Papineni2002-jv} between the correct output and the output the model produces for a given perturbation. This goal function only assumes that the model output a prediction as a string. Given this design pattern, the \texttt{MinimizeBleuScore} goal function can be applied to attack any sequence-to-sequence model. Similar goal functions can be designed for other output formats, like classification models or sentence taggers.

\section{Common functions for text inputs with \texttt{AttackedText}}

Across TextAttack modules, some functionality is required over and over again. Many transformations want to split text inputs into a list of words. Many constraints require part-of-speech tagging. We want to avoid repeating code in too many places, and also to set a standard as to which tokenization, part-of-speech tagger, etc. is used. 

Therefore, with the exception of models (which take string inputs), TextAttack modules operate on \texttt{AttackedText} objects – not vanilla Python strings. The AttackedText contains string functionality that performs word replacement, prepares text to input to the model, prints inputs along with their column names, and manages attack-specific context attributes.

It is relatively common for NLP libraries to provide some base class that provides additional functionality to what are essentially enhanced string objects. For example, flair \cite{flair-akbik2018coling} performs text-level operations on a \texttt{Sentence} class. TextAttack follows a similar strategy and stores each text input as an \texttt{AttackedText} object.

\subsection{Everything is a single string}

A single input may consist of multiple strings. TextAttack transformations apply string-level transformations to inputs – for example, reordering words, or replacing a single word with its synonym. Most transformations are defined in the attack papers to operate on a single string-input. For multi-input classification tasks, adversarial attacks often just choose a single input on which to operate, like the hypothesis in the case of entailment \cite{TextFooler-Jin2019-re}.

TextAttack enables such single-string transformations and constraints without restricting itself to single-input tasks. Transformations and constraints assume the input is a single string. The AttackedText contains a property (\texttt{AttackedText.text}) that joins all text inputs with a space in between. This text value is passed to each transformation \& constraint, and then broken up again by column. 

\begin{table}[]
\scalebox{1.0}{
\begin{tabular}{|c|c|c|}
\hline
\textbf{Attack} & \textbf{Queries} & \textbf{Cache hits} \\ \hline
\citet{alzantot2018generating}   &  1029    & 736 \\ \hline
\citet {pso-zang-etal-2020-word} & 3745 &  3080  \\ \hline
\end{tabular}
}
\caption{``Queries'' stands for average number of queries to victim model to attack one sample, while ``cache hits'' represents the average number of times a query has resulted in a hit to the model output cache. Each cache hit saves a query to the model, so more cache hits indicates a higher performance boost due to caching.}
\label{table:caching}
\end{table}

\section{Improving Performance}
\paragraph{Model inference memoization}
Adversarial attacks in NLP spend most of their time on the GPU. For each text input, the attack must obtain the model's output, as well as the output of any models used to apply certain linguistic constraints, like a sentence encoder to ensure semantic similarity between adversarial example and the original text. Upon further examination, many of these model inferences appear over and over again during the attack process. For example, the attack needs to compute the model's score for an input that has already been seen. Some population-based stochastic search methods, like the genetic algorithm of \citet{alzantot2018generating}, may revisit the same input multiple times during the search process, which increases the number of redundant computations.

TextAttack caches model outputs to avoid redundant computations. This is done using a least-recently-used (LRU) function cache. Since outputs are generally small, TextAttack can maintain a very large LRU cache for each purpose without using an excessive amount of memory. In some cases, this high-level caching can cause a significant performance increase. We experimented with attacking 100 samples for \texttt{BERT-base} model \cite{devlin2018BERT} trained on SST-2 dataset \cite{sst-2-dataset} using methods proposed by \citet{alzantot2018generating} and \citet{pso-zang-etal-2020-word}. Table \ref{table:caching} shows that in both cases, significant number of queries to the victim model result in hits to the model output cache, helping us save time by avoiding unnecessary computations.

\paragraph{Multiprocessing strategy}
Efficient use of GPUs is critical for any deep learning job. If a GPU is available, TextAttack attacks typically use it for victim model inference and for inference on any models required for constraints. These inference times are the main bottleneck for many attacks. On systems with multiple GPUs, running attacks on samples sequentially results in use of only one GPU. We provide multiprocessing feature with the \texttt{--parallel} flag to instead runs attacks in parallel.

TextAttack parallel mode works by starting a new attack worker process for each GPU. Each worker takes dataset samples off of an in-queue, runs an attack on a single sample, puts the attack result on an out-queue, and repeats, until the in-queue is empty. An additional non-GPU worker works to print attack results as they appear on the out queue.

This multiprocessing paradigm is quite simple, and works nicely with various current deep learning packages. Other libraries that face similar single-GPU-intensive workloads could employ this pattern to parallelize many GPUs. In the future, the additional help of a distributed computing interface like MPI could allow an attack to be run across multiple machines as well.

\section{Enabling use across different operating systems and devices}

\paragraph{Operating system compatibility}
Different operating systems follow different filesystem conventions. Specifying full file paths explicitly is almost never a good idea. Instead, prefer using absolute paths. TextAttack uses absolute paths and combines filenames using Python's \texttt{os.path.join} utility function. This enables file manipulation on any system (not just Unix).

\paragraph{GPU Hubris}

Current deep learning frameworks allow explicit device placement of tensors – choosing whether a given tensor is on CPU or a specific GPU. It is easy to design specifically for your system: putting each tensor explicitly on the GPU where it belongs. However, this hurts cross-system compatibility: the code is now only able to run on systems with GPUs. TextAttack checks to see if CUDA is available before putting tensors on the GPU, and puts them on the CPU otherwise. This allows the library to run on machines without GPUs.


\section{Conclusion}

Writing an excellent, well-documented library that is easy to install and run is a good way to get researchers interested in a research topic as it lowers the barriers to entry. Moreover, a well-structured, extendable design empowers newcomers to make their contributions to the field. We hope that our lessons from developing TextAttack will help others create user-friendly open-source NLP libraries.

\newpage
\section*{Acknowledgments}

Thanks to all the TextAttack contributors who helped us solve these tough problems– including Eli Lifland, Jake Grigsby, Di Jin, Kevin Ivey, Alan Zheng, and others. Thanks also to Robin Jia and Paul Michel who provided invaluable feedback toward the development and design of TextAttack.

\bibliographystyle{acl_natbib}
\bibliography{emnlp2020}

\end{document}